# Trajectory eigenmodes of an orbiting wave source

E. Fort[1] and Y. Couder[2]

[1] *Institut Langevin, ESPCI ParisTech and Université Paris Diderot, CNRS UMR 7587, 10 rue Vauquelin, 75 231 Paris cedex 05, France.*
[2] *Matières et Systèmes Complexes, Université Paris 7 Denis Diderot, CNRS - UMR 7057, Bâtiment Condorcet, 10 rue Alice Domon et Léonie Duquet, 75205 Paris cedex 13, France.*

**Abstract**

Resonances usually result from wave superpositions in cavities where they are due to the wave spatio-temporal folding imposed by the boundaries. These energy accumulations are the signature of the cavity eigenmodes. Here we study a situation in which wave superposition results from the motion of a source emitting sustained overlapping waves. It is found that resonances can be produced in an unbounded space, the boundary conditions being now defined by the trajectory. When periodic trajectories are investigated, it is found that for a discrete subset of orbits, resonant wave modes are excited. Trajectory eigenmodes thus emerge. These modes have three attributes. Their associated resonant wave fields are the Fourier transform of the source's trajectory. They are non-radiative and they satisfy the perimeter Bohr-Sommerfeld quantization rule.

In quantum physics, wave-particle duality makes eigenmodes a possible characteristic of single objects. This is in sound contrast with classical physics where an external cavity is needed to fold waves to produce resonances. However, with the introduction of "walkers", wave-particle duality is no longer a characteristic of the sole quantum world [1-5]. These macroscopic objects, formed by a bouncing droplet guided by the wave it generates on a fluid interface, exhibit several phenomena similar to those characterizing the quantum world. They include probabilistic phenomena (single particle diffraction and interference [2], tunnel-like effect [3]…) as well as quantization behaviors (discrete Landau-like orbits [4] or Zeeman-like effect [5] …). The appearance of duality in this classical object suggests that it could be a generic property of a whole family of systems. Here we imagine a inertial version of such a system. We show that resonant modes emerge for specific orbits satisfying the Bohr-Sommerfeld (BS) relation.

A detailed introduction is needed to explain the origin of this model system and its path-memory driven dynamics. In the experiment, the trajectory of a single droplet is determined by its association to an individual wave-field playing the role of a pilot wave. This is reminiscent of the model introduced by de Broglie at an early stage of quantum mechanics in which a pilot-wave guides a single particle in the physical space [7,8]. The wave driven guidance of the particle requires that a part of the wave field remains at the particle location. This was achieved in de Broglie theory at the expense of causality. In his model, the pilot wave is a standing wave, a superposition of a retarded and an advanced wave [9]. The former can be thought of as emitted by the particle, but the latter is paradoxical as it converges onto the particle coming from the future. This same counter-intuitive hypothesis is needed in the Wheeler-Feynman absorber theory [10]. In contrast, the macroscopic walker's experiment provides a solution that restores causality in pilot wave theories. In this case, the droplet emits a divergent propagative front exciting quasi-sustained standing waves through the Faraday parametric instability (see Fig. 1a). The waves are sustained for the characteristic memory time τ which depends on the distance to the Faraday instability threshold [6]. In this system, the retro-propagating waves play the role of advanced waves. Since they are triggered by the outgoing propagating wave front emitted in the past they could be said to come "from the future of the past". In the Huygens-Fresnel theory using secondary sources, these sources would be isotropic and thus generate a counter-propagating wave front. It is interesting to note that this is opposite to standard wave propagation in which the retro-propagation is annihilated using dipolar sources.

The global wave-field driving the droplet results from the interference of these sustained standing waves generated at each bounce along the droplet's trajectory (see Fig. 1b). Hence the walker's wave encodes information on the droplet's past trajectory [6]. The resulting path-memory gives the walker a spatio-temporal non-locality at the origin of the quantum-like behaviors.

A moving walker generates standing waves motionless in the fluid frame of reference. Here, in order to restore a kind of Galilean invariance, we eliminate the specific role of the fluid reference frame. We postulate that the generated waves move in the droplet's inertial frame at the time of emission. In contrast with the experimental *hydrodynamic walkers* we will call these hypothesized objects, *inertial walkers*.

The space-time evolutions of hydrodynamic and inertial walkers are compared schematically in Fig. 2. The wave fields associated with a motion at constant velocity are compared in (a) and (b). In the experiment, because of the displacement of the emitter, the sources emitted at each bounce overlap, resulting in a complex interference pattern [6]. The extension of the wave-field is controlled by memory. In the case of an inertial walker, the standing waves generated at each bounce move with the particle. Hence these in-phase waves superpose to form an isotropic circular wave. A kind of Galilean invariance for the walker, as a whole, is thus restored.

Note that, in the experimental situation, the asymmetry of the wave-field produces a local slope that provides self-propulsion and compensates damping. In the inertial case, for a constant speed, the slope is zero. This would be consistent with a Newtonian frictionless dynamics. The inertial walker can thus be considered as a dissipation-free limit of the hydrodynamic walker. The existence of the memory has no effect for an inertial motion.

We now focus on circular motions. Figures 2c and 2d show schematically the two situations. In the hydrodynamic case, the centers of the standing circular waves generated at each bounce are evenly distributed along the droplet's orbit. The resulting wave-field, observed in the experiment was described in ref. [4]. Conversely, for an inertial walker, the circular wave packet generated at a given bounce drifts with the walker's velocity at the time of its emission. Its center appears as a secondary source ejected tangentially out of the orbit. The locus of these virtual sources forms a single armed Archimedean spiral (see Fig. 2d).

Let us recall the main experimental findings for orbiting hydrodynamics walkers. Circular orbits had been obtained using Coriolis force as the hydrodynamic analog of magnetic Lorentz force. In this situation where a transverse force is applied to the droplet, the memory has radical effects. The path-memory dynamics generates an additional "quantization" force. Only a discrete set of circular orbits is authorized [4]. The existence of a Landau-like quantization was demonstrated, with a scaling where the Faraday wavelength $\lambda_F$ plays the role of the de Broglie wavelength $\lambda_{dB}$ of the electron. However, if these two quantities are exactly identified to one another, the observed quantization differs from the semi-classical Bohr-Sommerfeld condition in that the diameters of the orbits are multiples of the wavelength $2R=n\lambda_F$, instead of the perimeters.

We now investigate inertial walkers and their associated wave-fields. The standing waves produced at each bounce are centered on the virtual sources distributed on a spiral. As in the experiment, they are in-phase Bessel functions, their rate of creation is equal to the Faraday period $T_F$, and their initial spacing is small compared to $\lambda_F$ (small walker velocities). We investigate long path-memory dynamics for which quantization behaviors emerge.

The numerical simulations of the global wave field reveal the emergence of a series of strong successive resonances when the imposed orbit radius $R$ is increasing continuously (see Fig. 3 and supplementary Video). Their periodicity follows exactly the BS quantization of the orbital perimeter. They can be labeled by their order $n$ where $n=2\pi R_n/\lambda_F=k_F R_n$. Each resonance is characterized by an azimuthal periodicity of order $n$ and exhibits radially a series of circular nodes. In contrast, out-of-resonance wave fields are disordered and of small amplitude (Fig. 3b). Resonances become narrower ($Q$ factor) as the time spent in orbiting motion is increased.

The whole structure rotates with the droplet and is invariant in its frame of reference. At each resonance the radial profiles can be fitted accurately with a Bessel function $J_n(k_F r)$ of the first type and order $n$ (see Fig. 4). The entire wave-field for the $n$th resonance can be fitted by the function $J_n(k_F r)\cos(n\theta+\phi_n)$ where $\phi_n$ is the relative angular shift between the global wave field and the position of the walker.

These results can be obtained analytically. Here, we give the detail of the calculation that derives the wave field structure at resonance. The calculations are performed within the following approximations: long memory time $\tau$ compared with rotation period along the orbit $2\pi/\Omega$ ($\tau\Omega \gg 1$), small walker's velocity $V_W$ compared with the Faraday wave phase velocity ($V_W \ll \lambda_F/T_F$) and limit of infinite front wave propagation velocity. The global wavefield in $\vec{r}$ at time $t_0=0$ is the sum of individual wave fields generated at each bounce of the form $J_0(k_F|\vec{r}-\vec{r}_l|)e^{-(t_0-t_l)/\tau}$ where $J_0$ is the zero order Bessel function of the first kind centered on the position $\vec{r}_l$ of the $l$th bounce in the past, that occurred at time $t=-lT_F$. At time $t_0=0$, the global wavefield is given by:

$$\varphi(\vec{r}) = \sum_{l=0}^{\infty} J_0(k_F|\vec{r}-\vec{r}_l|)e^{-lT_F/\tau}$$

This expression can be written using addition formula as follows:

$$\varphi(\vec{r}) = \sum_{l=0}^{\infty} \left( \sum_{n=-\infty}^{+\infty} J_n(k_F r)e^{in\theta} J_n(k_F r_l)e^{-in\theta_l} \right) e^{-lT_F/\tau}$$

Using the relation $J_{-n}(x)=(-1)^n J_n(x)$ and introducing the arbitrary angle $\phi$, we can write the global wave field in term of positive Bessel functions:

$$\varphi(\vec{r}) = A_0 J_0(k_F r) + \sum_{n=1}^{+\infty} J_n(k_F r)[A_n \cos(n\theta+\phi) + B_n \sin(n\theta+\phi)]$$

with 
$$\begin{cases} A_0 = \sum_{l=0}^{\infty} J_0(k_F r_l)e^{-lT_F/\tau} \\ A_n = 2\sum_{l=0}^{\infty} J_n(k_F r_l)\cos(n\theta_l+\phi)e^{-lT_F/\tau} \\ B_n = 2\sum_{l=0}^{\infty} J_n(k_F r_l)\sin(n\theta_l+\phi)e^{-lT_F/\tau} \end{cases}$$

For bounces occurring after one revolution of the walker around the orbit, *i.e.* for bouncing times satisfying $l > L \approx 2\pi/(T_F\Omega)$, the angular position of the $l$th bounce satisfies with a very good approximation: $\theta_l \approx \Omega T_F l$. Since the loci of the centers of the sources form an Archimedean spiral, the position of the $l$th bounce satisfies with a good approximation: $r_l \approx R\Omega T_F l$.

Hence, we can write:

$$\begin{cases} A_0 = \sum_{l=L}^{\infty} J_0(k_F R\Omega T_F l)e^{-lT_F/\tau} \\ A_n = 2\sum_{l=L}^{\infty} J_n(k_F R\Omega T_F l)\cos(n\Omega T_F l + \phi)e^{-lT_F/\tau} \\ B_n = 2\sum_{l=L}^{\infty} J_n(k_F R\Omega T_F l)\sin(n\Omega T_F l + \phi)e^{-lT_F/\tau} \end{cases}$$

$J_n(k_F R\Omega T_F l)$ are asymptotically periodic functions with angular frequencies equal to $k_F R\Omega T_F$. Hence the sum of its product with the sinusoidal function of frequency $n\Omega T_F$ is resonant only if both functions have the same frequency, *i.e.* for: $k_F R = n$, which is the analog of Bohr-Sommerfeld relation: $2\pi R = n\lambda_F$. Note that because of the oscillatory behavior of Bessel functions, $A_0$ is always negligible compared to the resonant coefficient $A_n$ or $B_n$. Through the adequate choice of the arbitrary angle $\phi$, it is possible to tune the non oscillatory component. Since $A_n$ and $B_n$ are in quadrature, $\phi$ can be set to a value $\phi_n$ which maximize $A_n$ and minimize $B_n$. We thus obtain: $\varphi(\vec{r}) \approx A_n J_n(k_F r)\cos(n\theta + \phi_n)$. $\phi_n$ depends *a priori* on the order $n$ of the orbit. Finite memory prevents the divergence of $A_n$ and thus of the global wave field. However, note that although the guidance phenomenon of the hydrodynamic walker is well understood, its non dissipative generalization is still to be resolved.

The resonant behavior is intriguing since there is no boundary forming a cavity. The emergence of the resonance is a direct consequence of the path-memory dynamics. The resonant wave-field is built-up by the relative spatial position of the inertial sources so that they interfere constructively. Inertial sources have spatially correlated phases and provide a collective mode for specific periodic trajectories. Thus the standard spatial folding needed for resonance comes from the periodicity of the particle trajectory, not from boundaries imposed to the wave.

The Bessel expansion $J_n(k_F r)\cos(n\theta)$ of $n$th order, is also the 2D Fourier transform of a circle of radius $k_F$ with an azimuthal phase modulation $\cos(n\theta)$. It is remarkable that at resonances, the global wave field is the Fourier transform of the walker's trajectory. Path-memory dynamics thus results in the surprising coexistence of the two reciprocal objects in the same physical space.

Bessel modes are known to be diffraction-free solutions of Helmholtz equation [11,12]. Their associated energy flux is azimuthal and they possess a discrete angular momentum proportional to their order. Hence, at resonances, the walker's wave field is also endowed with these unexpected quantization and non-radiative properties.

Path-memory relies on the idea that the elementary waves produced at each bounce are sustained for a certain time. However, more generally, the only requirement is an overlap of the emitted waves.

The resonant orbits of memory endowed sources incite the discussion of possible analogies with electrons that also satisfy BS quantization. The idea that a standing wave is associated with a moving electron is due to de Broglie [10]. However, if an analogy was to be found with inertial walkers, a variant of his hypotheses would have to be considered where the electron would be dressed with new memory attributes. The production of waves could originate from its internal degree of freedom, *i.e.* the spin. The classical analog of the spin being an orbiting charge [14], it could generate a stable in time Bessel substructure. If the electron itself orbits, the memory effect would result in the build-up of a global wave field. The latter would become resonant for those orbits that satisfy the BS relation. Furthermore, these Bessel resonances do not radiate, a property that Bohr had to postulate in his early atomic model [15]. Thus, this would define stationary orbits that are the eigenmodes of the electron.

**Acknowledgments**

We thank E. Bossy, J. Bush, R. Carminati, A. Eddi, M. Fink, M. Labousse, M. Miskin, J. Moukhtar and S. Perrard for fruitful discussions. We thank the AXA Research Fund and the French National Research Agency (project FREEFLOW and LABEX WIFI) for financial support.

**Figure captions**

**Figure 1.**

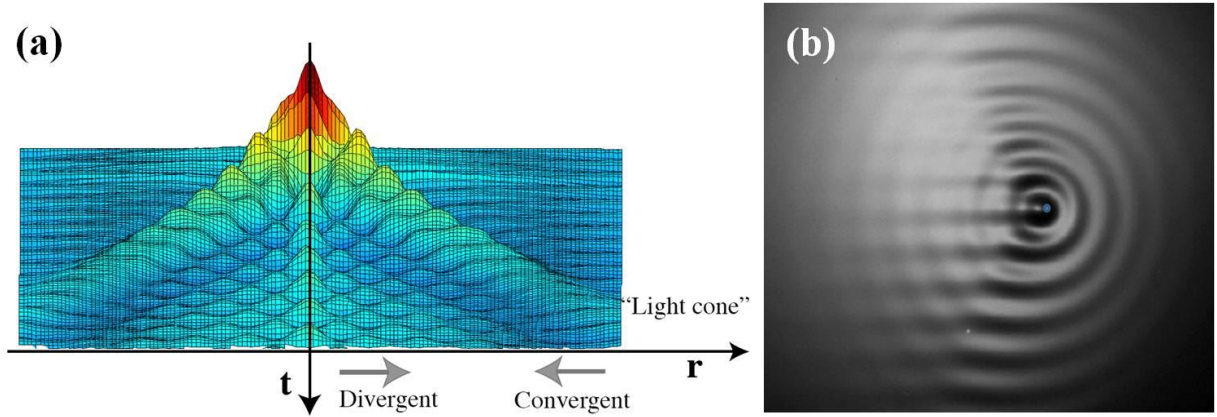

The hydrodynamic walker's wave field. (a) Spatio-temporal evolution of the surface profile generated by a *single* point-like disturbance of a vibrated bath close to the Faraday instability threshold. Time elapses from top to bottom. The outer envelope corresponds to the propagative capillary wave front. This disturbance is sufficient to trigger Faraday standing waves that damp out slowly. This wave contains a convergent "advanced-like" wave. The survival of the wave field at the initial disturbance point is the source of the memory effect. (b) Photograph of a walker moving at constant velocity. The wave field is the superposition of the sustained Faraday waves emitted at each bounce. The memory effect generates a Fresnel type of interference pattern.

**Figure 2.**

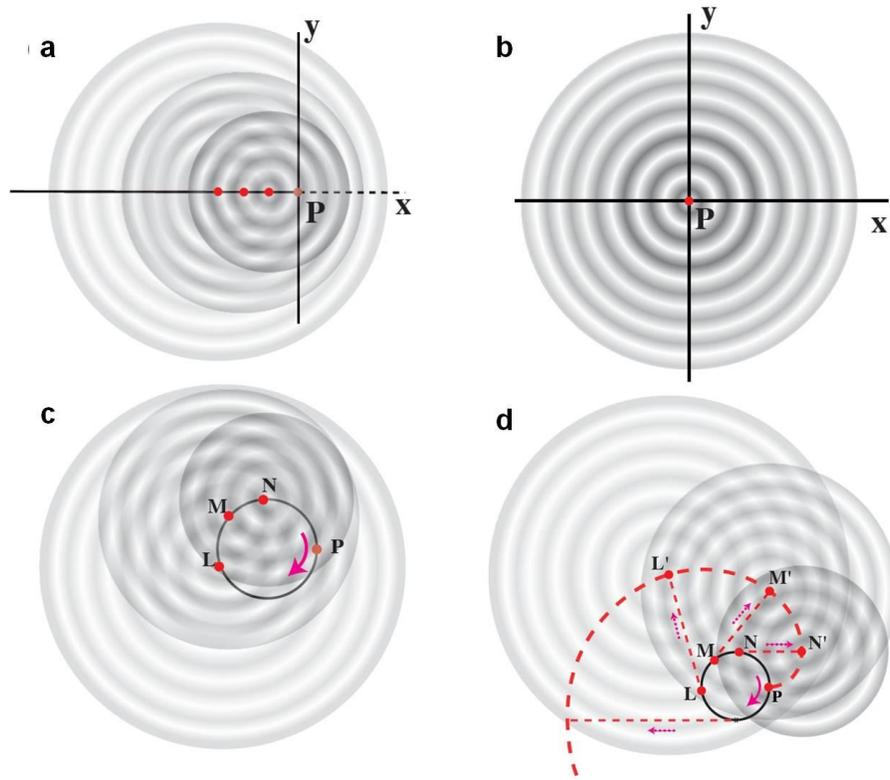

Schemes of the build-up of the global wave fields due to path-memory for a hydrodynamical Walker (HW) and *Inertial Walkers* (IW) in straight and circular motion. In all cases the wavefield results from a linear superposition of successive elementary Bessel wave-packets generated along the particle's trajectory. For HW, the wave-packets are fixed in the laboratory frame of reference while for IW, they move with the instantaneous speed of the particle at the time of emission. For the rectilinear trajectories, the waves generated by the HW form a "wake" with a complex interference pattern in **a** while with IW all wave packets remain centered around the particle in (b). For circular orbits, in the HW case, the locus of the centre of the secondary sources is the orbit itself in **c**. In the IW case, the wave packets unfold tangentially. Their locus is an Archimedean spiral in (d).

**Figure 3.**

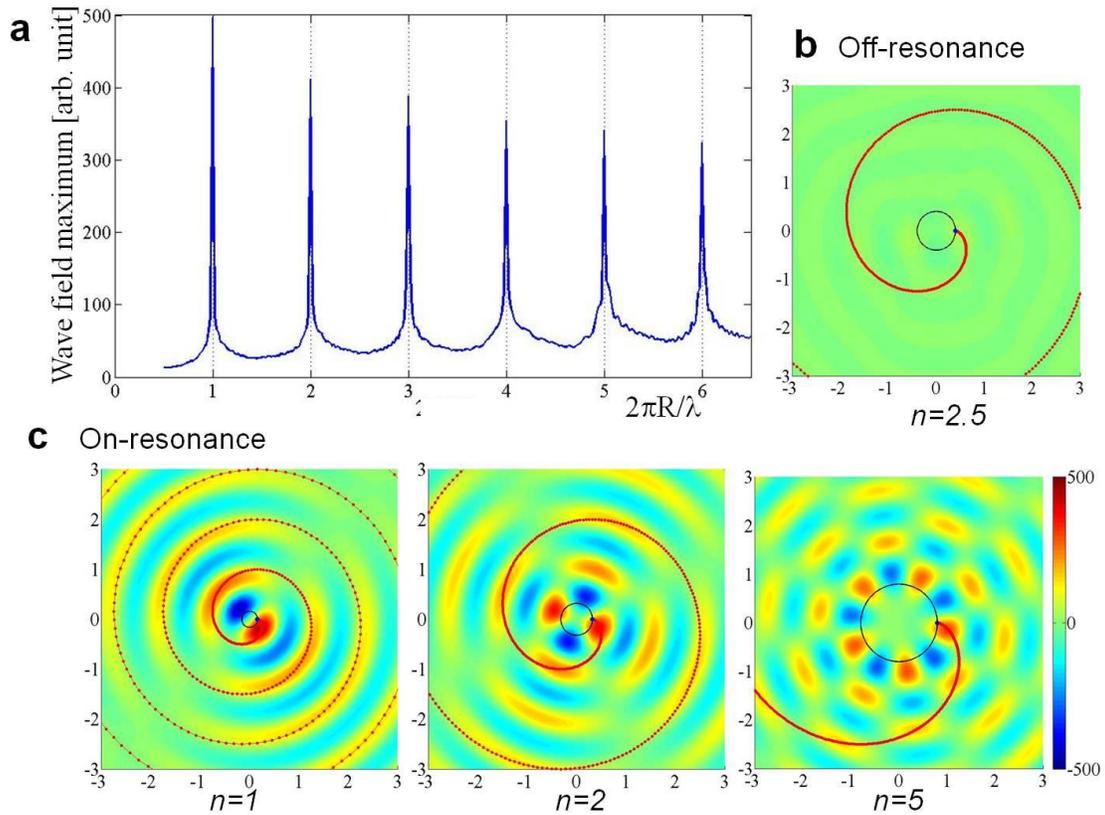

Wave field resonances. (a), Amplitude of the wave field maximum versus the normalized perimeter of the orbit $2\pi R/\lambda_F$. Sharp resonances satisfying a kind of Bohr-Sommerfeld relation $2\pi R=n\lambda_F$, $n$ being a non zero integer, are clearly visible. Examples of associated wave field in resonances (b) for $n=1$ and (c) for $n=5$. The color scale is the same for the two wave fields, the back circle is the walker's trajectory, the connected red dots are the centers of the inertial sources emitted in the past, placed on an Archimedean spiral, the blue dot is the current position of the walker. The simulations correspond to 10000 bounces of a walker moving at 1% of the Faraday phase velocity.

**Figure 4.**

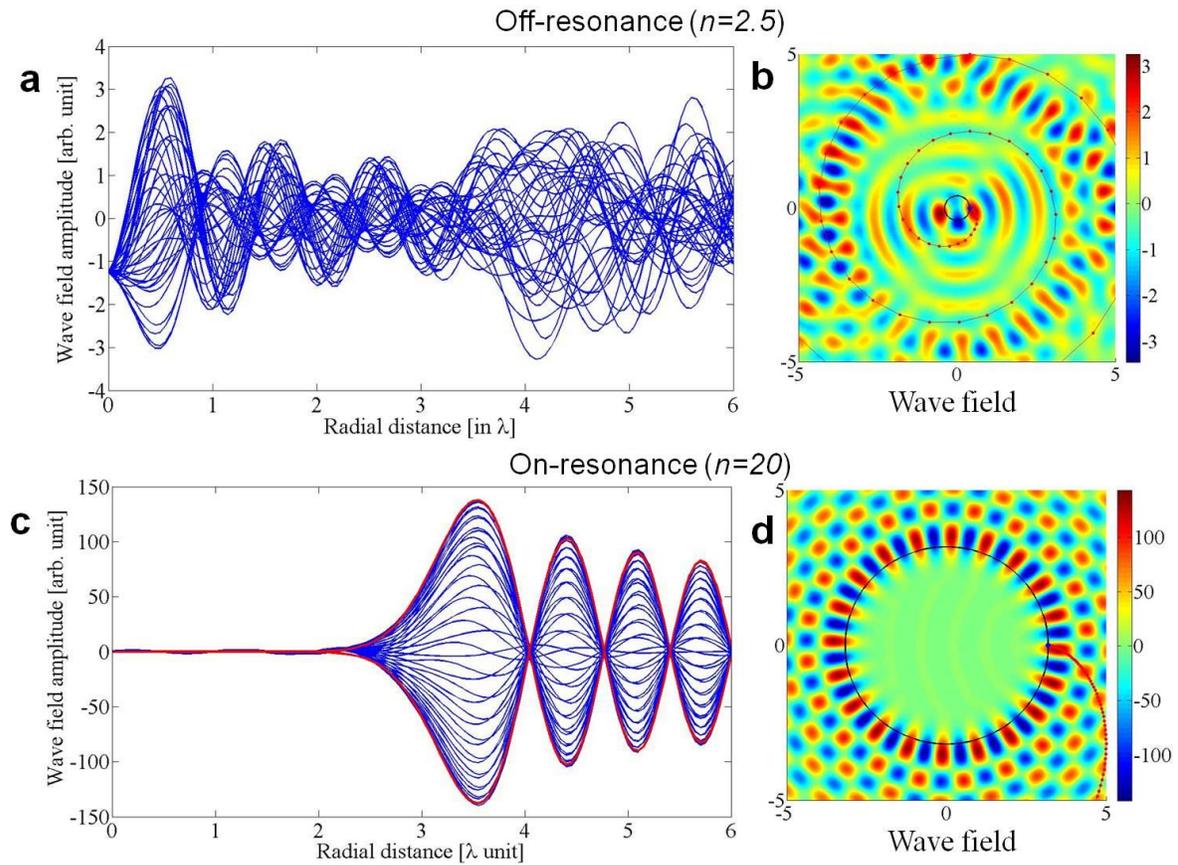

Wave field profiles. Radial profiles (blue curves) of the wave field along various directions and the associate wave field: off resonance for *n=2.5* in (a) and (b) respectively, in resonance for *n=20* in (c) and (d) respectively. Note the amplitude change of scale of about 100 folds between the off and in resonance wave fields and profiles. The envelope of the resonance profile for *n=20* is accurately fitted by a Bessel function $J_n(k_F r)$ of order *n* (red curve). The simulations correspond to 80000 bounces of a walker moving at 10% of the Faraday phase velocity.

**Supplementary multimedia**

The movie shows the evolution of the global wave field when increasing continuously the radius of the orbit $2\pi R/\lambda_F$. The orbit is represented as a black central circle. The position of the walker and of the centers of the emitted inertial Bessel waves are represented respectively as blue and red dots. The simulations correspond to 10000 bounces of a walker moving at 1% of the Faraday phase velocity.